\documentclass[pra,twocolumn,showpacs,superscriptaddress,10pt]{revtex4-1}
\usepackage[colorlinks,citecolor=blue,linkcolor=red,dvipdfm]{hyperref}
\usepackage{amsmath,amssymb,graphicx}
\usepackage{times}
\usepackage{float}

\begin{document}

\title{Electromagnetically induced moir\'{e} optical lattices in a coherent atomic gas}

\author{Zhiming Chen}

\affiliation{State Key Laboratory of Transient Optics and Photonics, Xi'an
Institute of Optics and Precision Mechanics of Chinese Academy of Sciences, Xi'an 710119, China}
\affiliation{School of Science, East China University of Technology, Nanchang 330013, China}
\affiliation{Collaborative Innovation Center of Light Manipulations and Applications, Shandong Normal University, Jinan 250358, China}

\author{Xiuye Liu}
\affiliation{State Key Laboratory of Transient Optics and Photonics, Xi'an
Institute of Optics and Precision Mechanics of Chinese Academy of Sciences, Xi'an 710119, China}
\affiliation{University of Chinese Academy of Sciences, Beijing 100049, China}

\author{Jianhua Zeng}
\email{\underline{zengjh@opt.ac.cn}}
\affiliation{State Key Laboratory of Transient Optics and Photonics, Xi'an
Institute of Optics and Precision Mechanics of Chinese Academy of Sciences, Xi'an 710119, China}
\affiliation{University of Chinese Academy of Sciences, Beijing 100049, China}
\date{\today}

\begin{abstract}
Electromagnetically induced optical (or photonic) lattices via atomic coherence in atomic ensembles have recently received great theoretical and experimental interest. We here conceive a way to generate electromagnetically induced moir\'{e} optical lattices---a twisted periodic pattern when two identical periodic patterns (lattices) are overlapped in a twisted angle ($\theta$)---in a three-level coherent atomic gas working under electromagnetically induced transparency. We show that, changing the twisted angle and relative strength between the two constitutive sublattices, the moir\'{e} Bloch bands that are extremely flattened can always appear, resembling the typical flat-band and moir\'{e} physics found in other contexts. Dynamics of light propagation in the induced periodic structures demonstrating the unique linear localization and delocalization properties are also revealed. Our scheme can be implemented in a Rubidium atomic medium, where the predicted moir\'{e} optical lattices and flattened bands are naturally observable.

\end{abstract}

\maketitle

\section{Introduction}{\label{Sec:1}}
Spatial periodic structures, particularly the man-made ones like waveguide arrays, photonic crystals and lattices, as well as optical lattices, have great applications in controlling the flow of light and matter waves because of their intriguing structural properties (e.g., partial and full photonic band gaps, symmetry-protected topological spectrum)~\cite{PCF,PC,OL-RMP,rev-light,rev-trapping,DarkGS,Frac-CQ2d,FOP-focusing,FOP-QD,DarkGS-Q,DarkGS-CQ}. The fabrications of such artificial periodic structures and investigations of the peculiar wave properties have been receiving great research attention in past years. Of particular interest in the optics and photonic communities are the two fabricated methods, direct femtosecond-laser writing technique and optically induced ones, with the former being widely used in solid materials where the induced photonic lattices have a permanent refractive index (and the optical and thermal stability of the laser machining method should be carefully processed) and the latter applies both to solid materials and gaseous media. The periodic structures manufactured by the latter method are aliased as electromagnetically induced gratings (EIGs), which are being extensively studied from both theoretical and experimental sides~\cite{EIG1,EIG2,EIG3,EIG4,EIG5,EIG6,EIG7,EIG8,EIG9,EIG10,EIG11,EIG12} in  recent years, owning to the highly tunable degree of freedom of the induced periodic structures enabled by external and real-time changeable environments for both hot atomic vapours under room temperature and ultracold atoms (like Bose-Einstein condensates) in the nano-Kevin regime.

Electromagnetically induced transparency (EIT) is a unique quantum interference in coherent atomic ensembles with multilevel electronic structures, where a strong control (light) field dresses the field-coupled states and then a weak probe field cannot feel the absorption, leading to the cancellation of strong absorption completely in the induced transparency's spectral region and thus making the atomic medium transparent~\cite{EIT-RMP}. Interesting properties and promising futuristic applications that are closely associated with the EIT in an all-optical way include (but are not limited to) coherent population trapping, greatly enhanced nonlinear susceptibility, steep dispersion, slow and fast light, shape-stable coupled excitations of light and matter (the so-called dark-state polaritons), light storage (including nonlinear wave localization), communications and computations in both classic and quantum regimes
~\cite{EIT-DSP,EIT-storage,EIT-applications,Dynamical-PBG,EIT-blockade,EIT-OL1d,EIT-OL2d,Bai,RW,HHK,XCH,ZLLL,CXLH,CBLHH}. Particularly worth mentioning is the fact that, the EIGs with tunable optical properties (lattice depth, periodicity, structural arrangement, etc.) under EIT regime are pushing towards the realization of those application targets~\cite{EIG1,EIG2,EIG3,EIG4,EIG5,EIG6,EIG7,EIG8,EIG9,EIG10,EIG11,EIG12}.

In very recent years, scientists are progressing toward the realization of novel periodic structures, and particularly moir\'{e} patterns~\cite{moire1,moire2}---two-dimensional (2D) twisted structures of two identical periodic structures overlapped in a twisted angle ($\theta$)---are entering the optics and photonics communities while their fabrication by means of EIGs in gaseous medium still remains blank. Here a realizable all-optically way depended on the aforementioned optically induced technique for fabricating the moir\'{e} optical lattices in a three-level $\Lambda$-type atomic system under EIT is conceived. Rich and interesting extremely flat bands of the underlying band-gap structures are discovered for the electromagnetically induced moir\'{e} optical lattices,  showing unique linear localization and delocalization moir\'{e} physics for light propagation as displayed in other 2D moir\'{e} structures reported elsewhere~\cite{moire1}.

\section{Theoretical model}
\begin{figure}
\centering
\includegraphics[scale=0.412]{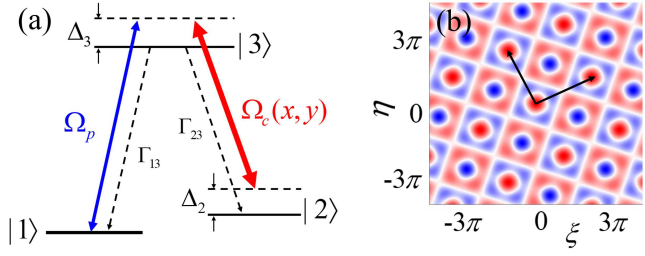}
\caption{(Color online) (a) Theoretical configuration of a $\Lambda$-type three-level atomic system that induces moir\'{e} optical lattices under EIT.  A weak probe field $\Omega_p$ couples the ground state $|1\rangle$ to the excite state $|3\rangle$, and a strong space-dependent control field $\Omega_c$ couples the metastable state $|2\rangle$ to the excite state $|3\rangle$. The spontaneous emission decay rates of transitions $\Gamma_{13,\,23}$ and detunings $\Delta_{2,\,3}$ are defined in the text. The combined effect of weak probe and strong control fields, $\Omega_p$ and $\Omega_c$, kindles the standard EIT regime. (b) Contour plot of the spatial periodic modulated control field $\Omega_c(x,y)$ that is taken as a moir\'{e} pattern (shaded blue, pattern minima; shaded red, pattern maxima) at $\epsilon_1=\epsilon_2 =1$ with rotation angle $\theta=\arctan(3/4)$. The black arrows denote the primitive vectors.}
\label{fig:1}
\end{figure}

{\label{Sec:2}}
Our fundamental theory relates to light propagation in a three-level $\Lambda$-type coherent atomic ensemble working in EIT regime, as described in Fig. \ref{fig:1}(a). A weak probe field with half-Rabi frequency $\Omega_p$ and center angular frequency $\omega_p$ dresses ground state $|1\rangle$ to excite state $|3\rangle$, and a very strong continuous-wave control field with half-Rabi frequency $\Omega_c$ and center angular frequency $\omega_c$, modulated periodically in spatial, connects metastable state $|2\rangle$ to excite state $|3\rangle$. The spontaneous emission decay rates of transitions $|3\rangle \rightarrow |1\rangle$ and $|3\rangle \rightarrow |2\rangle$ are represented by $\Gamma_{13}$ and $\Gamma_{23}$, and the detunings $\Delta_2=\omega_{p}-\omega_{c}-\omega_{21}$ and $\Delta_3=\omega_{p}-\omega_{31}$, here $\omega_{jl}=(E_j-E_l)/\hbar$ with $E_j$ being the eigen energy of state $|j\rangle$. To obtain the electromagnetically induced moir\'{e} optical lattices under the EIT condition, the control field (after simplification) is chosen as the periodic function of spatial coordinates $(x,\,y)$, i.e.,
\begin{eqnarray}\label{field}
&&\Omega_c(x,y)=\Omega_{c0}[1+f(x,y)].
\end{eqnarray}
Here $\Omega_{c0}$ is a constant describing the magnitude of the control field, and $f(x,y)=\epsilon_1(\cos^2 x+\cos^2 y)+\epsilon_2(\cos^2 x^\prime+\cos^2 y^\prime)$, $\epsilon_{1,\,2}>0$ being the modulation depth (amplitude) of the two optical lattices with the same periodicity $\pi$. The rotated $(x^\prime, y^\prime)$ plane at a rotating angle $\theta$ yields
\begin{equation}\label{eig}
\left(
\begin{array}{cccc}
 x^\prime\\
 y^\prime
\end{array}
\right )
=\left(
\begin{array}{cccc}
 \cos\theta & -\sin\theta\\
 \sin\theta & \cos\theta
\end{array}
\right )
\left(
\begin{array}{cccc}
 x\\
 y
\end{array}
\right ).
\end{equation}
For discussion, we define the strength contrast as $p=\epsilon_{1}/\epsilon_{2}$. The contour plot of the control field $\Omega_c(x,y)$  given by Eq. (\ref{field}) taken as moir\'{e} square lattice is displayed in Fig. \ref{fig:1}(b), showing a spatial twisted displacement compared to the conventional square lattice.

By adopting the standard Maxwell-Bloch equations that describe the propagation of probe field $\Omega_{p}$ , and after substitution as detailed in {\color{blue}Supplement Information}, we have the dimensionless 2D envelope equation
\begin{eqnarray}\label{EE}
&&i\frac{\partial u}{\partial
s}=-\frac{1}{2}\left(\frac{\partial^2 }{\partial \xi^2}+\frac{\partial^2}{\partial \eta^2}\right)u
+V_{\mathrm{OL}}(\mathbf{r})u,
\end{eqnarray}
where the dimensionless variables are defined as spatial coordinates $\mathbf{r}=(\xi,\,\eta)=(x,\,y)/R$, $u=\Omega_p/U_0$, and propagation distance $s=z/L_{\rm Diff}$ with typical diffraction length $L_{\rm Diff}=\omega_pR^2/c$. Here $U_0$ and $R$ are the typical Rabi frequency and beam radius of the probe field. The coefficient of the last term in Eq. (\ref{EE}) represents the induced moir\'{e} optical lattice potential with lattice depth $V_0$ [See {\color{blue}Supplement Information}]
\begin{equation}
V_{\mathrm{OL}}(\mathbf{r})=-\frac{V_0}{[1+f(\xi,\eta)]^2}.
\label{OL}
\end{equation}

{The theoretical model considered here can be realized in realistic physical systems. Specifically, the energy levels $|1\rangle$, $|2\rangle$, and $|3\rangle$ can be selected respectively as $5^{2}\textrm{S}_{1/2}(F=1)$, $5^{2}\textrm{S}_{1/2}(F=2)$, and $5^{2}\textrm{P}_{1/2}(F=2)$ states of $^{87}$Rb atoms tuned to D1-line transition~\cite{Steck}, and the decay rates are given by $\Gamma_2\simeq 1.0$ kHz, and $\Gamma_3\simeq 5.75$ MHz, and $|\textbf{p}_{13}|=2.54\times10^{-27}$ C cm. To achieve  the dimensionless 2D envelope equation (\ref{EE}), other parameters are chosen as $\mathcal{N}_a=3.69\times10^{10}\,V_0$ cm$^{-3}$ ($V_0$ is a real constant denoting lattice depth), $\Omega_{c0}=1.0\times10^{7}$ Hz, $R=36$ $\mu$m, $\Delta_1=0$, $\Delta_2=1.0\times 10^5$ Hz, and $\Delta_3=1.0\times 10^4$ Hz, and thus the typical diffraction length is $L_{\rm Diff}=1.0$ cm. Note that these parameters are used in all our calculations reported below.}

\begin{figure*}[httb]
\centering
\includegraphics[width=1.9\columnwidth]{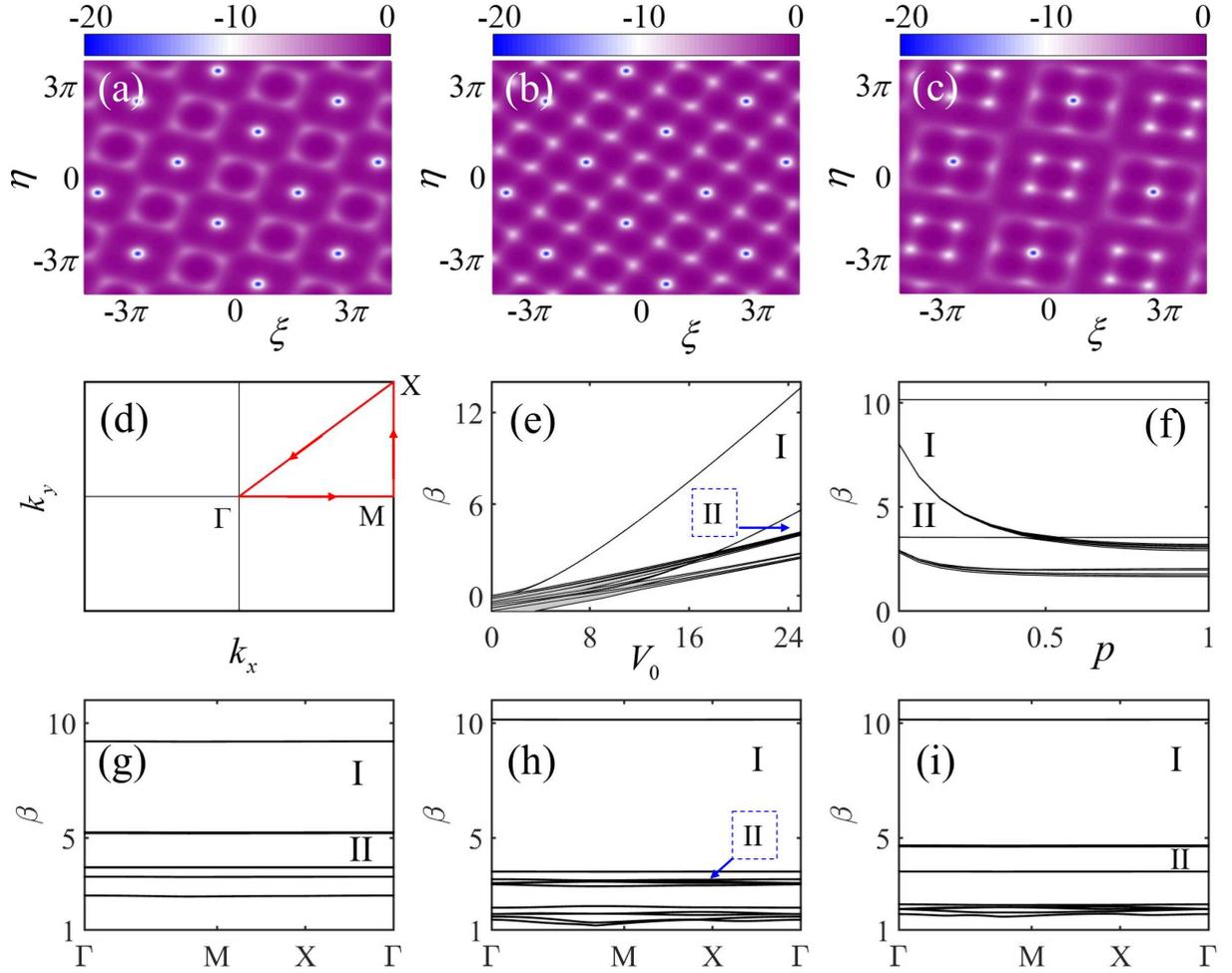}
\caption{Contour shapes of the electromagnetically induced moir\'{e} optical lattice potentials (shaded blue, potential  minima; shaded purple, potential maxima) with different parameters: (a) twisted angle $\theta=\arctan(3/4)$ and strength contrast $p=1$ ($\epsilon_1=\epsilon_2=1$), (b) $\theta=\arctan(3/4)$  and $p=0.25$ ($\epsilon_1=0.4$, $\epsilon_2=1.6$), (c) $\theta=\arctan(5/12)$ and strength contrast $p=1$. We have set $V_0=20$ for panels (a)-(c). (d) The first Brillouin zone of the  induced lattice potentials in the reciprocal space; signed are high symmetry points ($\Gamma$, M, X). The linear Bloch-wave spectra (expressed by propagation constant $\beta$) of the induced optical lattice at twisted angle $\theta=\arctan(3/4)$ with (e) varying $V_0$ and $p=1$, (f) different ratios $p$ and $V_0=20$. Band-gap structures of the lattices at $p=1$ and $V_0=20$ and under different twisted angles: (g) $\theta=\arctan(3/4)$ and (h) $\theta=\arctan(5/12)$. (i) Band-gap structure with $p=0.25$ and $V_0=20$ at $\theta=\arctan(3/4)$ . I and II in panels (e)-(i) represent the first and second finite band gaps, respectively.}
\label{fig:2}
\end{figure*}

\section{Numerical results}
We stress that once again the electromagnetically induced moir\'{e} optical lattices [given in Eq. (\ref{OL})] provide a rotational degree of freedom for the periodic structures, in contrast to those conventional periodic optical patterns formed also under the EIT condition. As pointed out elsewhere~\cite{moire0}, when Pythagorean angle is satisfied, $\theta=\arctan(a/b)$, $\cos\theta=a/c$ and natural numbers $(a, b, c)$ obey $a^2+b^2=c^2$, the moir\'{e} optical lattices Eq. (\ref{OL}) can be defined as periodic structures in the first Brillouin zone, and then the corresponding band gap structures can be easily obtained using the Bloch theory. To this end, the probe field $u$ can be written as $u=U(\mathbf{r}) \exp(i\beta s)$, where $\beta$ is the propagation constant, and the stationary envelope $U(\mathbf{r})$ can be sought as the form $U(\mathbf{r})=\phi_{\mathbf{k}}(\mathbf{r})\exp(i\mathbf{k}\cdot\mathbf{r})$, here the wave vectors $\mathbf{k}=(k_x, k_y)$ are confined to the  first reduced Brillouin zone of the moir\'{e} optical lattices, and $\phi_{\mathbf{k}}(\mathbf{r})=\phi_{\mathbf{k}}(\mathbf{r}+\mathbf{d})$ is a periodic Bloch function having the same periodicity $\mathbf{d}$ as the lattices. Then the dispersion relation of the 2D Bloch waves, $\beta(\mathbf{k})$, can be found by calculating the linear eigenvalue problem
\begin{eqnarray}\label{eigenvalue}
\left[\frac{1}{2}\left(\frac{\partial }{\partial \mathbf{r}}+i\mathbf{k}\right)^2
-V_{\mathrm{OL}}(\mathbf{r})\right]\phi_{\mathbf{k}}(\mathbf{r})=\beta(\mathbf{k})\phi_{\mathbf{k}}(\mathbf{r}).
\end{eqnarray}
{Numerically, we adopt the plane wave expansion method~\cite{PC} to calculate the eigenvalue problem  (\ref{eigenvalue}), periodic boundary condition is applied based on the Bloch theory. Such processing way is matured in optics context for processing photonic crystals, more details could be referred to Ref.~\cite{PC}.}

Figures \ref{fig:2}(a)$\sim$\ref{fig:2}(c) display the shapes of the induced moir\'{e} optical lattices under different Pythagorean angles $\theta$ and strength contrast as $p=\epsilon_{1}/\epsilon_{2}$ {with $V_0=20$.} To calculate the associated band gap structures, we have given the first Brillouin zone in the momentum (reciprocal) space $(k_x, k_y)$ in Fig. \ref{fig:2}(d), whose exact values can be tuned to portray the electromagnetically induced moir\'{e} lattice potentials with varying twisted angles $\theta$. The linear band-gap structures of the lattices, described as propagation constant $\beta$ versus lattice depth $V_0$ and versus strength contrast $p$   $(=\epsilon_{1}/\epsilon_{2})$, are respectively displayed in Figs. \ref{fig:2}(e) and \ref{fig:2}(f). It is observed from the former that the width of the first finite band gap widens and more higher band gaps emerge with an increase of $V_0$. At $V_0=20$ while increasing $p$, the first finite band gap widens while the second gap narrows, according to the latter. The rich spectra (and the variation rule with changing $V_0$ and $p$) of the moir\'{e} lattice potentials in Figs. \ref{fig:2}(e) and \ref{fig:2}(f) supplement and enrich the band-gap properties of the moir\'{e} patterns in other optics backgrounds~\cite{moire1,moire2,moire0}, implying the possibility of finding flat band feature of the induced lattices in our model.

At a defined $V_0$ and $p$, the underlying band-gap structures of the induced moir\'{e} lattice potentials change dramatically with different twisted angle $\theta$, comparing the Figs. \ref{fig:2}(g) and \ref{fig:2}(h); interestingly, there are a lot of flatten bands, which are a unique feature of moir\'{e} patterns~\cite{moire1,moire2,moire0}. One can see that at $\theta=\arctan(3/4)$, there exists a wide first band gap and the second gap is in a moderate width; keeping other conditions constant while setting $\theta=\arctan(5/12)$, a much wider first gap is created and the second gap shrinks greatly.  The strength contrast $p$ is also an important parameter that can change the band-gap structure of the induced moir\'{e} lattices, through changing it, the widths of the first finite band gap and the second one can be tuned almost by will, according to Fig. \ref{fig:2}(i) and as compared with Fig. \ref{fig:2}(g). As emphasized above, the electromagnetically induced periodic structures have a tunable advantage, since the power, structural arrangement, and dimensionality of the laser fields could be modulated based on the experimental requirements; in this respect, not only the moir\'{e} square lattices described here, but also the moir\'{e} aperiodic lattices and the hitherto unexplored spectrum property can also be envisaged, which will be the future direction.

\begin{figure}
\centering
\includegraphics[scale=0.42]{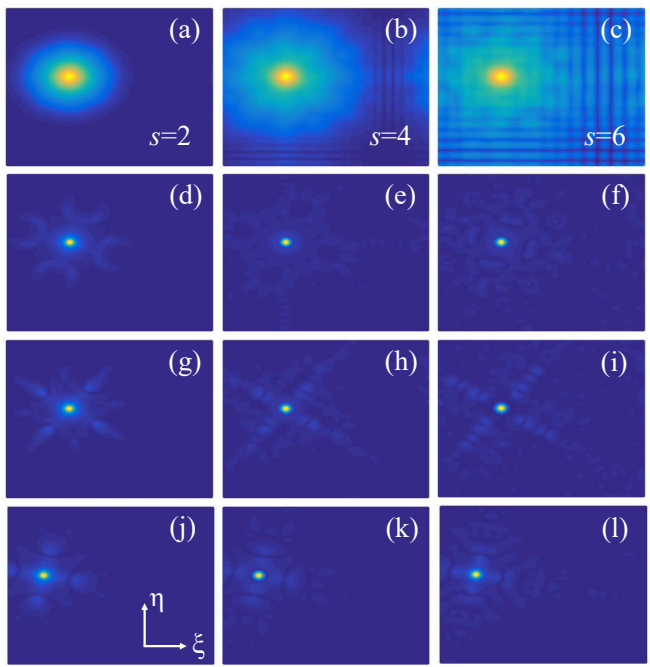}
\caption{Linear propagations of light (module, $|u|$) in the induced moir\'{e} optical lattices at twisted angle $\theta=\arctan(3/4)$ and different propagation distance $s$ {($s=z/L_{\rm Diff}$, as given below the Eq. (\ref{EE}) ) }displaying delocalization [(a)$\sim$(c)] and localization [(d)$\sim$(f), (g)$\sim$(i)] properties. Other parameters: (a)$\sim$(c) strength contrast $p=1$, lattice depth $V_0=1$; (d)$\sim$(f) $p=1$, $V_0=20$; (g)$\sim$(i) $p=0.25$, $V_0=20$. (j)$\sim$(l) show the linear propagations of light in the induced moir\'{e} optical lattices at twisted angle $\theta=\arctan(5/12)$ with $p=1$, $V_0=20$ for the light propagating to $s=2$, $s=4$, and $s=6$, respectively. {$\xi$, $\eta$ $\in [-12,\,12]$ for all panels}.}
\label{fig:3}
\end{figure}

The most recent theoretical predictions and experimental progresses of moir\'{e} lattices in optics community have proven the spectacular linear localization mechanism provided by the flat bands of the lattice~\cite{moire1,moire-nanolaser}. It is thus natural to ask whether such new localization paradigm or regime can be complied with in our electromagnetically induced moir\'{e} optical lattices. Our answer is a definitely yes. Figure \ref{fig:3} demonstrates such linear localization and delocalization properties of light propagation in the moir\'{e} optical lattices introduced here when it propagates to $s=2,\,4,\,6$, respectively. Under delocalization paradigm, light propagation in the moir\'{e} lattice undergoes the typical diffraction pattern along the propagation distance, as seen in Figs. \ref{fig:3}(a)$\sim$\ref{fig:3}(c) for such evolutions {at twisted angle $\theta=\arctan(3/4)$ with strength contrast $p=1$ and lattice depth $V_0=1$}; under the localization regime offered by the flat bands, both the shape and strength of the input light could be well conserved, as displayed by two typical examples in the second line and the third one of Fig. \ref{fig:3} {at twisted angle $\theta=\arctan(3/4)$ and $V_0=20$ with $p=1$ and $p=0.25$, respectively. It is seen that the localized light in Figs. \ref{fig:3}(g)$\sim$3(i) has more weak side patterns compared to that of Figs. \ref{fig:3}(d)$\sim$3(f), this difference is induced by the different configurations of the moir\'{e} optical lattices (as comparing the Figs. \ref{fig:2}(a) and \ref{fig:2}(b)), apparently, there is a stronger Bragg diffraction since the existence of several higher sublattices around the lattice minima for the case of $p=0.25$. We also demonstrate that the twisted angle of the moir\'{e} optical lattices has a great impact on the light propagation, figures \ref{fig:3}(j)$\sim$3(l) display the linear propagations of light at twisted angle $\theta=\arctan(5/12)$ with $p=1$ and $V_0=20$, which reveals that the shape and position of light change as the twisted angle changes  in the localization and delocalization mechanisms.} It is relevant to mention that the light evolutions in Fig. \ref{fig:3} are produced by means of both the fast-Fourier transform and fourth-order Runge-Kutta methods, and both methods can smoothly match. The initial light input we used is the 2D Gaussian beam with appropriate amplitude and waist.

\section{Conclusion}
Summarizing, we have conceived a way to generate 2D moir\'{e} lattice potentials in a coherent atomic ensemble using the optically (electromagnetically) induced method. We find that the lattice depth $V_0$,  twisted angle $\theta$, and strength contrast $p$ could have great influences on the shapes and the band-gap structures of the resulting moir\'{e} lattice potentials, emphasizing the flat band property. The extraordinary localization and delocalization abilities of the induced moir\'{e} lattice are also confirmed. Considering the fact that our theoretical analyses are based on the realistic atomic media with realizable parameters, the predicted electromagnetically induced moir\'{e} lattices and the associated moir\'{e} physics can be readily observed in experiments. Future research interests may be paid attention to the moir\'{e} aperiodic lattices at non-Pythagorean angles $\theta$ and in nonlinear situation where the nonlinear localization of light in atomic gases is yet to be explored.

\vspace{4mm}
\noindent\textbf{Electronic supplementary material}
Electronic supplementary materials are available in the online version of this article at https://
 and http:// and are accessible for authorized users.

\vspace{4mm}
\noindent\textbf{Acknowledgments}
This work was supported by the National Natural Science Foundation of China (11704066, 12074423, 12074063), and Jiangxi Provincial Natural Science Foundation (20202BABL211013).


{}

\end{document}